\documentclass[acmtog]{acmart}

\newenvironment{datamaterial}%
{ \vspace{-0.15cm}%
    \small\noindent{\bfseries Availability of Data and Material:}\par%
    \noindent\ignorespaces}%
{ \par\noindent%
\ignorespacesafterend }%

\usepackage{algorithmic}
\usepackage{graphicx}
\usepackage{textcomp}
\usepackage{colortbl}
\usepackage{hyperref}
\usepackage{acronym}
\usepackage{todonotes}
\usepackage{multirow}
\usepackage{amsmath}
\usepackage{rotating}
\usepackage{xurl}

\AtBeginDocument{%
  }

\setcopyright{acmlicensed}
\copyrightyear{2023}
\acmYear{2023}
\acmDOI{XXXXXXX.XXXXXXX}

\acmJournal{TOG}
\acmVolume{37}
\acmNumber{4}
\acmArticle{111}
\acmMonth{8}

\begin{document}

\title{Distributed Multi-objective Optimization in Cyber-Physical Energy Systems}


\author{Sanja Stark}
\email{sanja.stark@offis.de}
\affiliation{
  \institution{OFFIS}
  \streetaddress{Escherweg 2}
  \city{Oldenburg}
  \country{Germany}
  \postcode{26121}
}

\author{Emilie Frost}
\email{emilie.frost@offis.de}
\affiliation{
  \institution{OFFIS}
  \streetaddress{Escherweg 2}
  \city{Oldenburg}
  \country{Germany}
  \postcode{26121}
}

\author{Marvin Nebel-Wenner}
\email{Marvin.Nebel-Wenner@be-storaged.com}
\affiliation{
  \institution{OFFIS}
  \streetaddress{Escherweg 2}
  \city{Oldenburg}
  \country{Germany}
  \postcode{26121}
}


\begin{abstract}
Managing complex \ac{CPES} requires solving various optimization problems with multiple objectives and constraints. As distributed control architectures are becoming more popular in \ac{CPES} for certain tasks due to their flexibility, robustness, and privacy protection, multi-objective optimization must also be distributed. For this purpose, we present MO-COHDA, a fully distributed, agent-based algorithm, for solving multi-objective optimization problems of \ac{CPES}. MO-COHDA allows an easy and flexible adaptation to different use cases and integration of custom functionality. To evaluate the effectiveness of MO-COHDA, we compare it to a central NSGA-2 algorithm using multi-objective benchmark functions from the ZDT problem suite. The results show that MO-COHDA can approximate the reference front of the benchmark problems well and is suitable for solving multi-objective optimization problems. In addition, an example use case of scheduling a group of generation units while optimizing three different objectives was evaluated to show how MO-COHDA can be easily applied to real-world optimization problems in \ac{CPES}.
\end{abstract}



\received{01 December 2023}

\maketitle
\begin{datamaterial}
The scenario used for the evaluation is available at: \url{https://github.com/Digitalized-Energy-Systems/MOO-CPES/releases/tag/Distributed_Multi-objective_Optimization_in_Cyber-Physical_Energy_Systems}. The resulting data can be found at: \url{https://zenodo.org/records/10245786}.
\end{datamaterial}
\section{Introduction}
\acresetall
\label{sec:introduction}

A significant number of \ac{DER} are introduced into the power system as a result of the energy transition. Many of the \ac{DER} are based on renewable resources, such as wind and solar energy. In the future, these will account for the majority of the electricity supply while the generation from conventional power plants will decline. As a result, power generation will no longer be purely demand-driven but also depend on the weather, season, and time of day. At the same time, the growing number of electric vehicles, heat pumps, and storage systems can change consumption patterns. In particular, distribution grids require greater automation to effectively incorporate \ac{DER} and manage the overall increasing system complexity, which calls for advanced \ac{ICT}. \ac{ICT} subsequently becomes an essential and integral component of the energy system, giving rise to a \ac{CPES}.

The efficient and safe planning and operation of \ac{CPES} can be described by various optimization problems with multiple objectives and constraints. Exemplary problems are managing energy systems on microgrid \cite{karimi2020optimal, khezri2020review} or household level \cite{mayer2020environmental}, optimal placement and sizing of \ac{DER} \cite{ramli2018optimal, nowdeh2019fuzzy} or system restoration \cite{almoghathawi2019resilience, stark2021your}. The objectives include minimizing environmental effects or $CO_2$ emissions, minimizing cost, minimizing power loss, and maximizing stability or reliability. 

Due to potential conflicts between these objectives, one optimal solution usually does not exist but rather a set of trade-off or \textit{Pareto-optimal} solutions, where the performance of one objective cannot be increased without reducing the other. 
Finding a Pareto front can be accomplished by population-based metaheuristics such as evolutionary or swarm-based methods. These algorithms search different regions of the solution space simultaneously and, with few modifications, can find a set of multiple non-dominated solutions in a single run \cite{konak2006multi}. Many different approaches from both of these algorithm classes have already been used in the context of energy system optimization, for example, particle swarm \cite{naderi2021novel}, slime mold \cite{khunkitti2021multi}, glow-worm \cite{salkuti2019optimal} or grey wolf \cite{wu2020novel} optimization from the swarm-intelligence group and NSGA-2 \cite{wang2020multi, ren2019multi, ruiming2019multi, sharma2020coordination} and MOGA \cite{ullah2021multi} from the evolutionary group. 

All these approaches use a central solver in which all necessary information is assumed to be available in one place. 
However, in the case of \ac{CPES}, distributed control architectures have become more popular for certain tasks. In this case, multiple control elements have the joint responsibility for deciding on control actions to fulfill defined objectives \cite{han2017taxonomy}.  
A distributed control allows plug-and-play of \ac{DER}, battery storages, or loads while at the same time being robust due to no single-point-of-failure \cite{espina2020distributed}. Therefore, distributed architectures offer more flexibility and robustness compared with centralized ones \cite{ren2021multi}. Moreover, distributed approaches allow for preserving the privacy of \ac{DER} and loads by storing technical details locally and only exchanging necessary information with other entities. This is particularly critical as it is common in \ac{CPES} for some information to be only available distributed rather than collected at a central point.

To perform optimization processes with distributed architectures, a \ac{MAS} is a suitable choice \cite{ren2021multi}. \ac{MAS}-based optimization in \ac{CPES} has not only been studied in the context of microgrid management \cite{karimi2020optimal, khezri2020review} but also, for example for the electricity market \cite{luo2018distributed} or power system restoration \cite{yang2021multiagent}. 
We therefore extended the work from Bremer et al. in \cite{bremer2019towards} and developed a fully distributed, agent-based algorithm to solve multi-objective optimization problems of \ac{CPES}. Due to the diversity of optimization problems in \ac{CPES}, we have ensured the extensibility and flexibility of the algorithm to allow easy adaptation to any problem at hand. Our goal is to provide a basic framework that allows to fully exploit the advantages of an agent-based multi-objective optimization -- e.g. having different agents with different solving strategies, which can also be dynamically adapted during the optimization and for the specific problem, e.g., taking into account the computational power each agent has.

This paper has the following contributions:
\begin{enumerate}
    \item Introducing \textit{MO-COHDA}: a \textbf{fully distributed agent-based multi-objective optimization algorithm}, that is capable of solving \textbf{distributed optimization problems with a potentially large number of decision variables} where each entity only has limited control of one decision variable.
    \item Evaluating the effectiveness of MO-COHDA by \textbf{comparing it to \ac{NSGA-2} using multi-objective benchmark functions} from the \ac{ZDT} problem suite.
    \item Showing with \textbf{an example use case for scheduling a group of generation units with regards to three different objectives} how MO-COHDA can be easily applied to real-life optimization problems in \ac{CPES}. 
\end{enumerate}

The remainder of the paper is organized as follows. \autoref{sec:related-work} gives an overview of related work in the context of distributed multi-objective optimization. \autoref{sec:mo-cohda} describes the proposed multi-objective algorithm. \autoref{sec:evaluationbenchmark} presents the evaluation of the algorithm with benchmark problems and \autoref{sec:evaluationcpes} with a \ac{CPES} problem. Finally, \autoref{sec:conclusion} gives a short summary of the results and describes possible future work.

\section{Related Work}
\label{sec:related-work}
Several approaches in the literature deal with multi-objective optimization in \ac{CPES}. \ac{CPES} are inherently decentralized, as the power is created primarily by small generating units close to the customers instead of central large power plants. For the coordinated optimization of hybrid AC/DC distribution networks, three categories are distinguished in \cite{wang2021decentralized}, which can be transferred to optimization problems in \ac{CPES}: First, in \textit{centralized optimization}, data is collected to perform decisions centrally. Second, in \textit{distributed optimization}, multiple instances coordinate to reach a collective decision, and each instance controls its object. Finally, \textit{decentralized optimization} is a state between centralized and distributed control. Since the distinction between centralized and distributed optimization is sufficient for our use cases, we divided the related work into centralized and distributed multi-objective optimization.


\subsection{Centralized Multi-objective Optimization}
Adiou et al. present a multi-objective optimization for the operation of decentralized multi-energy systems \cite{adihou2019multi}. The objectives considered are the optimal design, maximizing the renewable energy rate, and minimizing costs. The authors use a  genetic algorithm for their approach.
Additionally, Murray et al. deal with decentralized multi-energy systems, where the focus is Power-to-Mobility \cite{MURRAY2020117792}. The approach minimizes the costs and the life-cycle emissions of buildings and vehicles. A mixed integer linear program is used. 

Wang et al. consider optimizing a decentralized integrated energy system in northern rural areas \cite{WANG20223063}. Two objectives are regarded: the economy (including different kinds of costs) and environmental protection (minimizing pollutants). The optimization is done with \ac{NSGA-2}.

Using central optimization requires all the necessary information to be centrally available. This is not always feasible in CPES due to the large number of optimization variables as well as data privacy reasons. Therefore, a distributed algorithm for the optimization is suitable, as it does not rely on global information \cite{ren2021multi}.


\subsection{Distributed Multi-objective Optimization}
A distributed multi-objective optimization is done by Ren et al. \cite{ren2021multi}. The authors develop a distributed multi-objective algorithm for ice storage systems, combining particle swarm optimization with a differential evolution algorithm. The optimization goals are minimizing energy consumption, operation cost, and energy loss. In this approach, different nodes, called computing process nodes, take over the optimization task while exchanging information with other nodes. The nodes in this example calculate their local knowledge, make decisions, and then exchange information with others. Thus, the optimization does not rely on global information.
Another example is the work from Wang et al. in \cite{wang2021decentralized}. Using the alternating direction multiplier method, they perform a coordinated multi-objective optimization regarding operational costs, voltage deviation, and network losses in Hybrid AC/DC Flexible Distribution Networks.
The authors present their multi-objective optimization algorithm and discuss the advantages but do not compare the effectiveness with benchmark functions or the suitability for other applications by properties such as flexibility and expandability.

Regarding optimizations in the power system, \ac{COHDA} \cite{hinrichs2017distributed} is well suited because it is fully distributed. \ac{COHDA} has multiple advantages relevant to the usage in a power system, as it ensures convergence and termination. The heuristic is implemented in a \ac{MAS}, as it is well suited for this due to its distributive nature. An often considered use case is the scheduling of \ac{DER} \cite{hinrichs2017distributed, niesse2016controlled, hinrichs2013distributed}, but \ac{COHDA} has already been used for various use cases and with multiple extensions and adaptations in each case \cite{stark2021your, hinrichs2015self, holly2021dynamic, bremer2017agent}.
The heuristic considers privacy constraints by keeping local information, such as information about different \ac{DER} units, private to the distributed agents. Only the best decision is communicated to other agents during the optimization process.
\ac{COHDA} was already considered by Schrage et al. using multiple criteria \cite{schrage2023multi}.
The authors designed an evolutionary algorithm to generate optimized schedules concerning multiple criteria, integrating the local objective.
Additionally, Bremer et al. consider \ac{COHDA} for multi-objective optimization while implementing an S-metric selection algorithm into it \cite{bremer2019towards}. This allows fully distributed agent-based multi-objective scheduling for energy resources. The approach presented in this paper is based on the one proposed by Bremer et al. and extends it. Furthermore, we applied the optimization of multiple global targets, while Bremer et al. considered only local targets. Our extension is explicitly explained in \autoref{sec:mo-cohda}.

\section{MO-COHDA}
\label{sec:mo-cohda}
In this work, we extended \ac{COHDA} to apply it to multi-objective problems. We will refer to this version as MO-COHDA. In the following, we will briefly describe the mechanisms of \ac{COHDA} and then present the adaptions we introduced in MO-COHDA.

\subsection{Preliminary Work: COHDA and basic MO-COHDA}
The heuristic used in this paper is called \ac{COHDA}. It was introduced by Hinrichs and Sonnenschein in \cite{hinrichs2017distributed}. 
\ac{COHDA} is a fully distributed optimization heuristic that uses self-organization mechanisms for the optimization of a common target.

The key concept of the heuristic is an asynchronous and iterative best-response behavior of distributed agents.
Every single agent has a working memory that is exchanged with other agents. The working memory contains the global target function, the most up-to-date information about the decisions of all agents in the system, and a solution candidate for the optimization problem.
Once an agent receives a message containing the working memory of another agent, its algorithmic approach includes three steps:
\begin{enumerate}
    \item \textbf{\textit{Perceive}}: The agent updates information about the current decisions of other agents. The agent replaces its own solution candidate in case the candidate its neighbor suggests contains more elements or yields a better rating.
    \item \textbf{\textit{Decide}}: The agent then searches for its best solution considering its local constraints, the information about the current system state of other agents, and the global target. If the resulting system state yields a better rating regarding the global target than the current solution candidate, a new solution candidate is created, replacing the old one.
    \item \textbf{\textit{Act}}: If any component of the working memory has been modified, the agent sends its working memory to its neighbors.
\end{enumerate}

Bremer and Lehnhoff presented the first version of a multi-objective \ac{COHDA} approach for Pareto fronts in \cite{bremer2019towards}. In their work, they adapted \ac{COHDA} in several ways based on the concept of the S-Metric Selection Evolutionary Multi-Objective Algorithm (SMS-EMOA) \cite{beume2007sms}:
\begin{itemize}
    \item \textbf{Solution Representation}: Instead of one solution, each agent stores a Pareto front, including several different solutions (so-called \textit{individuals}) within its working memory.
    \item \textbf{Solution Quality Assessment}: In order to compare the quality of two solution candidates (i.e. two Pareto fronts approximations) they calculate the achieved hypercube volume -- or hypervolume --, a concept first introduced by Zitzler et. al in \cite{zitzler2004indicator}. The hypervolume measures the volume of the space between the approximated Pareto front and a chosen reference point that should be dominated by all Pareto-optimal solutions, thereby describing the space dominated by the front. A larger hypervolume indicates closer convergence towards the true Pareto front. For a graphical representation of the hypervolume in a 2D space, see \autoref{apdx:ref_point}.
    \item \textbf{Generation of new Solution Candidates}: Following the approach of evolutionary algorithms, new individuals are created from the old ones using \textit{selection}, \textit{mutation} and \textit{crossover} operators. For selection, individuals are randomly selected and then mutated by adding a Gaussian delta to one of the agent's decision variables. Furthermore, as a crossover operator, a uniform crossover is applied. Upon that, the agent performs a fast, non-dominated sort and removes the worst individual. This procedure is repeated for several iterations. Decoder functions, which abstract from individual capabilities as introduced in \cite{bremer2018hybridizing}, were used to respect individual constraints.
\end{itemize}

\subsection{Adaptions in MO-COHDA}
In this work, we further extend this approach by integrating the following changes:
\begin{itemize}
    \item \textbf{Fixed reference point:} To calculate the hypervolume for comparing two solution candidates, a reference point has to be selected, which should be dominated by all Pareto-optimal solutions. This is necessary to ensure the same comparison characteristics for each agent. 
    In SMS-EMOA, this reference point is defined using the vector of the current worst objective values increased by 1.0 \cite{beume2007sms}. 
    Since the current worst objective values can change, the reference point is usually recalculated for each iteration. However, differing reference points may interfere with the convergence process of MO-COHDA. In the \textit{decide} step, a new solution candidate is only created when it receives a better rating regarding the global target. Therefore, \ac{COHDA} converges when no agent can find a better configuration and all agents have the same solution candidate. Since agents act asynchronously and can receive messages in any order, any agent must reach the same decision regarding which candidate is the best to guarantee this convergence, no matter in which order the solution candidates are compared (transitivity). This transitivity is not given with a dynamic reference point, which can result in deteriorative cycles \cite{berghammer2012convergence}. See \autoref{apdx:ref_point} for a simple example of this behavior. To prevent this, we made it obligatory to choose the reference point initially at the start of the optimization process. The drawback of a fixed reference point is that the worst values for each objective need to be known beforehand. 
    \item \textbf{Flexible integration of local and global constraints}: Our approach provides a flexible interface to integrate the local constraints of the agents. For example, the resulting local flexibility could be represented as a discrete set of possibilities, or it can be defined as a continuous range of feasible values.
    Generally, any representation is supported as long as it is possible (a) to get an initial feasible decision without any further inputs and (b) for the mutation algorithm to integrate the given representation of flexibility.
    Global constraints have to be communicated to all agents along with the global target functions. The concrete implementation of constraints, e.g., not allowing infeasible solutions altogether or the use of a penalty function, is also flexible. Some constraints in \ac{CPES} use cases might require more detailed knowledge about other agents like their location, for example, if the total power produced by units in a certain area should not exceed a defined threshold. In this case, this information should be prepared so that only the selected agents can interpret it, to continue to guarantee data protection and privacy.
    
    \item \textbf{Flexible way of generating new solution candidates}: Regarding the way new solution candidates are generated (the \textit{decide} step within \ac{COHDA}), we offer a flexible basic structure that allows different ways to create new individuals. We distinguish between two parts within the process: \textit{pick} and \textit{mutate}. The \textit{pick} function is used to select any number of individuals from the current system state any number of times, which will be taken as input for the mutation (for example, pick one or pick all individuals). The \textit{mutate} function will now generate new individuals. It will receive the local constraints of the agent as well as the formerly chosen individuals. Again, the concrete way the mutation is implemented is flexible. However, only the decision variables of the local agent can be mutated, and the local constraints must always be respected. The result of the \textit{pick} function is a set of new individuals.
    Finally, a non-dominated sort is executed, and the worst individuals are removed until the initial number of individuals is reached (\textit{reduce}). This procedure is repeated for a predefined number of iterations. 
    \item \textbf{Minimal change for updates of other agents}: We introduce the possibility of setting a margin regarding the minimum improvement in hypervolume necessary to update other agents. If a new candidate whose hypervolume was not significantly improved (not above the selected amount) was found, no update to other agents would be sent, and the candidate would be discarded.
\end{itemize}

\begin{figure}
\centering
\includegraphics[scale=0.6]{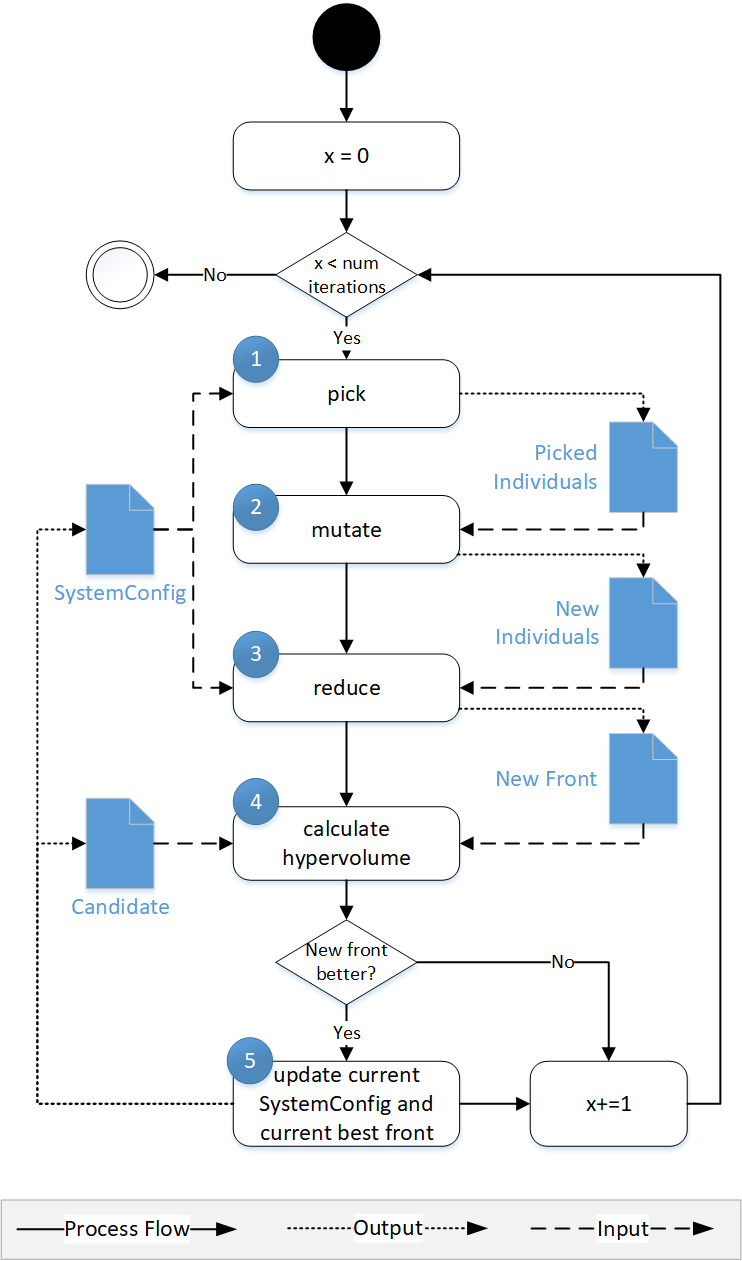}
\caption{Process of one iteration within \textit{decide}}
\label{fig:activity_diagram}
\Description[An iterative process is shown with the 5 steps described in the text]{The activity diagram shows an iterative process, starting with x=0. If x is smaller than the defined number of iterations, the process starts. It follows the five steps that are described in the text, pick, mutate, reduce, calculate hypervolume and update. For each step it is shown, which data objects are used as input and which data objects are created as output. Pick takes the SystemConfig as an input and creates Picked Individuals as an output. Mutate takes the picked individuals as input and creates new individuals as output. Reduce uses the SystemConfig and new individuals as input and creates new front as output. Calculate hypervolume takes the candidate and the new front as input. If the new front is better, update takes it as input and creates updated candidate and sysconfig as output. x is increased by one and the process starts over again. It is finished when x is equal to the number of iterations.}
\end{figure}

The extensions lead to the following process in \textit{decide}, as depicted in \autoref{fig:activity_diagram} and listed in the following. 
\begin{enumerate}
    \item \textit{Pick}: In each iteration, the individuals to be mutated are chosen from all solution points. 
    \item \textit{Mutate}: New individuals are generated by mutating the picked individuals using the selected mutation function. 
    \item \textit{Reduce}: The set of new individuals and the original individuals from the SystemConfig is reduced to the initial number of solution points, creating a new front.
    \item \textit{Calculate Hypervolume}: The hypervolume of the new front is calculated and compared with the hypervolume of the solution candidate. If it is not higher, the new front is discarded, and the next iteration begins with Step (1).
    \item \textit{Update} current SystemConfig and current best front: If the new front has a higher hypervolume, it replaces the old candidate, and the agent also updates his SystemConfig accordingly. The next iteration considers the updated information, beginning with Step (1). 
\end{enumerate}

In our implementation, solutions are always represented as "schedules", since this is a common representation for energy system scenarios where a set of \ac{DER}, loads, battery storages, and other flexible elements are coordinated over a period of time. One schedule is a time series of power values with one power value per time interval. However, the implementation can also be used for other representations. For example, if every agent controls just one variable in an optimization problem, it is possible to have a schedule with a single interval. This way, our implementation offers the flexibility and adaptability to multiple use cases. This also includes tuning the other parameters that are part of MO-COHDA, which are summarized in the following:
\begin{itemize}
    \item Number of solution points
    \item \textit{Pick} function
    \item \textit{Mutate} function
    \item Minimal change
    \item Number of local iterations
\end{itemize}

\subsection{Contributions of MO-COHDA}
MO-COHDA extends and improves the work from Bremer and Lehnhoff \cite{bremer2019towards} in several aspects. While still using the core functionality of the initial approach, MO-COHDA was explicitly developed to create a flexible and expandable, fully distributed multi-objective optimization heuristic that allows for a flexible definition of the agent's internal optimization strategies. All parts of the algorithm, including integration of objectives, constraints, decision variable format, flexibility representation, and optimization aspects, such as pick and mutate functions, can easily be adapted to test the best performance for a given problem. By introducing minimal change as an adjustable parameter, the convergence speed of MO-COHDA can also be influenced in one more aspect.

All the parameters can be set for each agent individually, allowing the testing of different optimization strategies in one optimization process. Overall, this allows more detailed testing of optimal parameter settings of the fully distributed multi-objective optimization concept for \ac{CPES} and addresses one of the points mentioned by Bremer and Lehnhoff as future work, namely the design of the optimization parameters. It should be noted that any performance improvement of MO-COHDA over the work from Bremer and Lehnhoff would only come from different parameter settings -- using the exact pick and mutate functions that Bremer and Lehnhoff used, the performance would most likely still be the same, as the core functionality of the SMS-EMOA was not changed.

Finally, the MO-COHDA algorithm is open source and comes with several options for parameter settings and use-case examples to make it easy to use. This includes, for example, a set of different mutate functions, ranging from a random mutation to a more advanced local optimization using NSGA-2. As MO-COHDA is
implemented within the multi-agent framework mango \cite{schrage2023mango}, it can easily be combined with other functionalities offered in mango, such as the termination detection or coalition formation.

\section{Evaluation - Benchmark Problem}
\label{sec:evaluationbenchmark}
To compare the performance of MO-COHDA with a central \ac{NSGA-2} solver, well-known MO benchmark problems and their respective Pareto fronts -- in the following called \textit{reference fronts} -- are used. While it would also be possible to represent an optimization problem with a single optimization variable in MO-COHDA -- by having each agent controlling only a part of this variable's search space -- we chose only to use benchmark functions with multiple optimization variables for comparison. This way, each agent is only responsible for one variable, which best reflects the usual use case for distributed optimization. 

In this Section, we introduce the chosen benchmark functions, describe the experimental setup in \autoref{subsec:benchmark_setup}, and discuss the evaluation results in \autoref{subsec:benchmark_results}.

\subsection{Optimization Problems}
\label{subsec:benchmark_problem}
We chose three functions from the \ac{ZDT} problem suite for evaluating MO-COHDA as they are well-known benchmark functions for MO optimization algorithms and fulfill our requirement for considering multiple variables \cite{zitzler2000comparison}. All problems are based on the same construction process and aim to minimize two objectives:
\begin{equation}
\begin{split}
    \text{min } f_1(x) & \\
    \text{min } f_2(x) & = g(x) h(f_1(x), g(x))
\end{split}
\end{equation}

The following equations depict the chosen problems, namely \ac{ZDT}1, \ac{ZDT}2, and \ac{ZDT}3. While the Pareto-optimal front from \ac{ZDT}1 is convex, the front from \ac{ZDT}2 is non-convex, and \ac{ZDT}3 has five disconnected fronts. The functions $f_1(x)$ and $g(x)$ are identical for all three problems: 

\begin{equation}
    \begin{split}
        f_1(x) & = x_1 \\ 
        g(x) & = 1 + \frac{9}{n-1} \sum_{i=2}^n x_1\\
    \end{split}
\end{equation}

The difference between the three problems lies in different $h(f_1,g)$. 

\begin{equation}
\tag{\ac{ZDT}1}
        h(f_1,g) = 1 - \sqrt{f_1/g}
\end{equation}

\begin{equation}
\tag{\ac{ZDT}2}
        h(f_1,g) = 1 - (f_1/g)^2
\end{equation}

\begin{equation}
\tag{\ac{ZDT}3}
        h(f_1,g) = 1 - \sqrt{f_1/g} - (f_1/g) \text{ sin}(10\pi f_1)
\end{equation}

For all three problems \begin{math} 0 \leq x_i \leq 1 \text{and } i = 1,...,n\end{math} applies, with n=30, resulting in 30 variables that need to be optimised. To solve these problems in MO-COHDA, each variable is represented by one agent that knows the allowed range between $0$ and $1$.

\subsection{Experimental Setup}
\label{subsec:benchmark_setup}
Here we present an example of a parameter setting that has led to good results in the solution of the benchmark functions. The decision variables were distributed among the agents to optimize \ac{ZDT}1-3 in a distributed way. Each agent is responsible for exactly one variable, leading to a total number of 30 agents. We represent the variables as one schedule with 30 entries, where each agent can only change one variable. At the beginning of the optimization, the agents do not know the values for any of the other agents' variables. Hence, they must make assumptions for the unknown variables, as the objectives can only be evaluated with all 30 values. Typically, an agent would assume 0 in this case. However, since the optimum for all \ac{ZDT} functions is \begin{math}0 \leq x_1 \leq 1 \text{ and } x_i=0 \text{ for } i=2,…,n\end{math}, assuming 0 would yield a very good initial performance, possibly distorting the results. Instead, agents assume the value 1 for all unknown variables to start from bad initial solutions. Another option would be to circumvent this issue altogether by having the agents wait with the actual optimization until they know an initial choice of each of the other agents.

The optimization process is now the following: When an agent receives its first \ac{COHDA} message, it chooses a random variable at its position in the schedule within its allowed range (anything between 0 and 1). The agent then updates its neighbors with only its own value changed. After this initial participation, an agent proceeds with the normal \textit{decide} process described in \autoref{fig:activity_diagram}. In the \textit{pick} function, all solution points are chosen to be mutated, with the \textit{mutation} function working as follows: for each picked solution point, the agent decreases the value by subtracting a random value (minimum to 0) and increases it by adding a random value (maximum to 1), thereby creating two new solution points. The range for the random value is set to $0.4-0.6$ for all three benchmark functions. 

A Pareto front with 25 solution points is created, and the reference point is set to $(1.1, 6.9)$, considering the worst possible values for $f_1$ and $f_2$ in all \ac{ZDT} functions are $1.0$ and $6.8$, respectively. The number of iterations is set to 1, as it seems due to the communication between the agents, which triggers regular calls of \textit{pick}- and \textit{mutate}, the optimization process is efficient even with just one iteration per \textit{decide}. 
The only parameter not identical for all three functions is minimal change, which is set to $0.0001$ for \ac{ZDT}1 and $0.001$ for \ac{ZDT}2,3 to achieve the best results. 

To evaluate the performance of our MO-COHDA implementation, we compare our results with the results of a central approach. We use the \ac{NSGA-2} implementation from the pymoo library and create Pareto fronts with 25 solution points to allow comparison \cite{blank2020pymoo}. All experiments for MO-COHDA and \ac{NSGA-2} are repeated 100 times. Furthermore, we consider the reference front for the benchmark problems.

\subsection{Results \& Discussion}
\label{subsec:benchmark_results}
\begin{figure}
\centering
\includegraphics[width=\linewidth]{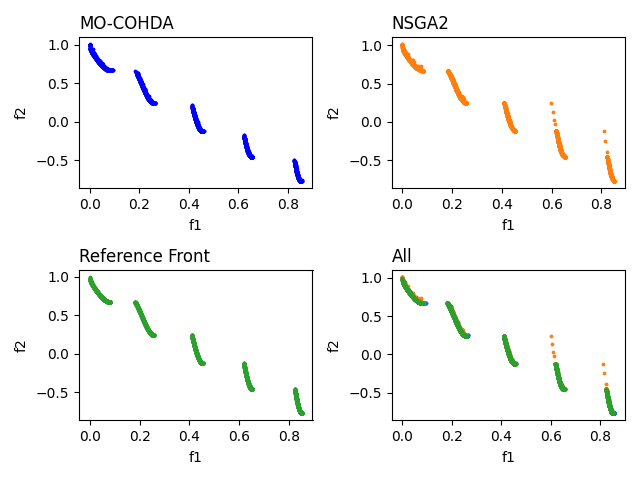}
\caption{Aggregated front of 100 runs each for MO-COHDA and Central approach compared with reference front for \ac{ZDT}3}
\label{fig:zdt3}
\Description[Four images show four graphs for MO-COHDA, NSGA-2, Reference front and All together. The fronts shown in all the pictures are nearly the same, except for a few outliers in the NSGA-2 front.]{The image is distributed in four graphs, each showing the front for ZDT3 created by MO-COHDA (top left), NSGA2 (top right), reference front (bottom left) and all fronts combined (bottom right). The reference front shows the true front for ZDT3 which consists of five disconnected fronts that start at the top left corner of the graph around (0,1) and continue until the bottom right corner of the graph around (0.85, -0.1). The fronts for MO-COHDA and NSGA-2 look nearly identical to the reference front, only NSGA-2 has a few outliers in some of the disconnected fronts. These are especially visible in the picture showing all fronts together, where the reference front does not cover all of the points found by NSGA-2.}
\end{figure}

\autoref{fig:zdt3} shows the aggregated fronts of 100 runs with 25 solution points each for both MO-COHDA and \ac{NSGA-2} for the \ac{ZDT}3 problem. For the results of \ac{ZDT}1 and \ac{ZDT}2 see \autoref{apdx:benchmark}. 
\autoref{fig:zdt3} shows that the two fronts are very similar and that the MO-COHDA implementation approximates the reference front very well.

To have a closer look at the results, the hypervolume and standard deviation for the simulation results are presented in \autoref{tab:benchmark_results}, for \ac{ZDT}1, \ac{ZDT}2, \ac{ZDT}3, considering the results of MO-COHDA, the centralized implementation, and the reference front. The larger the hypervolume, the better the performance of the Pareto front. 

\begin{table}[t]
\caption{Comparing Hypervolume mean and standard deviation of 100 runs for MO-COHDA and Central approach (rounded to 5 decimal places)}
\begin{tabular}{lccc}
\toprule
      Problem             &  MO-COHDA & Central & Reference\\
                       \midrule
\ac{ZDT}1 &   $7.236 \pm 0.000$     &  $7.230 \pm 0.002$      &  $7.256$ \\
\hline
\ac{ZDT}2 &    $6.904 \pm 0.001 $      &     $6.898 \pm 0.002$   &  $6.923$\\
\hline
\ac{ZDT}3 &    $7.699 \pm 0.001 $     &   $7.682 \pm 0.029$     &  $7.712$\\
                       \bottomrule
\end{tabular}
\label{tab:benchmark_results}
\end{table}

It is recognizable that the MO-COHDA implementation shows excellent results. The values for the hypervolume for all three benchmark functions are very close to the values of the reference front (maximum deviations of 0.02). The standard deviation of the solutions is also very low. If we look at the results of the central solution for comparison, it can be seen that these are similar to, and in some cases, slightly worse than, those of the MO-COHDA implementation. It should be noted, however, that the choice of a suitable \textit{mutate} function can greatly influence the goodness of the results. While moving in random steps towards the border values of the allowed range seems to be a good approach for the \ac{ZDT} problems, it might not give as good results for other problems. 
Overall, the results of the benchmark problems show that the implementation of MO-COHDA is suitable for use in multi-objective optimization.

\section{Evaluation - CPES Problem}
\label{sec:evaluationcpes}
With the second evaluation, we want to show the functionality of MO-COHDA for problems specific to \ac{CPES}. We have chosen an exemplary multi-objective problem from the energy domain and used MO-COHDA to solve it. \autoref{subsec:problem_cpes} describes the chosen problem, and \autoref{subsec:problem_setup} describes the implementation of this problem in MO-COHDA with two different parameter settings to introduce the various parameters. Finally, \autoref{subsec:results_cpes} presents the results from both settings.

\subsection{Optimization Problem}
\label{subsec:problem_cpes}

The goal of the exemplary problem is to optimize the schedules of a set of $n$ \ac{CHP} units and $m$ wind power plants, considering three different and potentially conflicting objectives. A schedule $\boldsymbol{s^c}$ of \ac{CHP} unit $i$ is described by $\boldsymbol{s_i^c} = (s_{i1}^c,...,s_{il}^c)$ with $l$ defining the length of the schedule and therefore the number of time intervals that this schedule covers. The schedule $\boldsymbol{s^w}$ of wind power plant $j$ is described by $\boldsymbol{s_j^w} = (s_{j1}^w,...,s_{jl}^w)$ respectively. Each element of a schedule defines the active power value of the respective \ac{CHP} or wind power plant for this time interval. The sum of all schedules of both, \ac{CHP} units and wind power plants, describes the cluster schedule $\boldsymbol{u}$ with \begin{math} \boldsymbol{u} = \sum_{i=1}^m s_i^c + \sum_{j=1}^n s_j^w\end{math}. 
The first objective is to minimize the difference between the cluster schedule $\boldsymbol{u}$ and a target schedule $\boldsymbol{t} = (t_1,...t_l)$. The target schedule could, for example, be defined by restrictions from the grid or a given target of a virtual power plant. To achieve a single performance value, the sum of all differences per interval of the schedule is considered using the Manhattan distance:

\begin{equation}
    \text{min } f_1 =  \sum_{k=1}^l \left | \left | \boldsymbol{t} - \boldsymbol{u} \right | \right|_k 
\end{equation}

The second objective is to minimize the share of produced power in the cluster schedule causing emissions. In this example, only the \ac{CHP} plants cause emissions, so only their contribution to the total produced power is considered. Again, the individual values for each schedule interval are summed up to one total emissions share:

\begin{equation}
    \text{min } f_2 = \sum_{x=1}^l \frac{\sum_{i=1}^m s_{ix}^c}{u_x} 
\end{equation}

Finally, the third objective aims at minimizing the uncertainty of the resulting cluster schedule. This means minimizing the share of power produced by units where the real available power can deviate from the forecasted power, in this case, the wind power plants. We assume that the uncertainties increase with time, as schedule intervals further in the future are more challenging to forecast and can deviate more from the estimated value. Therefore, we include a weight vector $\boldsymbol{w} = (w_1,...w_l)$ that assigns an uncertainty weight to each schedule interval with \begin{math} w_i = \left(100 / \frac{l (l+1)}{2}\right) \cdot i\end{math}, creating a linear increase in weights with all weights summing up to $100$. 
The objective is to minimize the sum of all weighted uncertain power shares over all schedule intervals: 

\begin{equation}
    \text{min } f_3 =  \sum_{y=1}^l \frac{w_y \sum_{j=1}^n s_{yj}^w}{u_y} 
\end{equation}

For optimal comparison of the objectives, all performance values are normalized between 0 and 1, using the $f_{min}$ and $f_{max}$ values. While $f_{min} = 0$ for all objectives, $f_{max}$ deviates. $f_{1max}$ is either the sum of all values in the target schedule or the sum of the maximum power values produced in each timestep by all units, depending on which value is larger. $f_{2max} = l$, meaning that all of the produced power causes emissions in each schedule interval. Finally, $f_{3max} = 100$, which occurs when in each schedule interval, the complete produced power is considered uncertain.

\subsection{Experimental Setup}
\label{subsec:problem_setup}

We created a scenario with 15 \ac{CHP} units (7x200kW, 8x400kW) and 15 wind power plants (2x 200kW, 3x 250kW, 3x300kW, 3x350kW, 3x400kW), which have to plan their power schedules for the next 6 hours, meaning they have to choose an active power value for 24 15-minute-intervals. Each \ac{CHP} has ten different possible schedules to choose from: one "off"-schedule consisting only of 0 values and nine schedules which were created using the $generate\_schedules()$-method from the pysimmods \ac{CHP} model\footnote{https://gitlab.com/midas-mosaik/pysimmods}. The wind power plants do not have schedules but a maximum power value they can produce in each interval. We used the data from 50Hertz\footnote{https://www.50hertz.com/de/Transparenz/Kennzahlen/ErzeugungEinspeisung/
EinspeisungausWindenergie}, starting 30.01.2022 at 12:30:00 pm and continuing over the following 6 hours. The power values were then scaled down to match the sizes of the wind power plants. Usually, wind turbine power control aims to produce the maximum amount of power while maintaining safe operating conditions, which can be achieved using various stall- and pitch-control techniques \cite{apata2020overview}. However, for the \ac{CPES} example, we also allow for a reduction of power output for wind plants: A wind power plant can create a power schedule by choosing any integer value between the maximum and 0 for each interval. This complements the limited flexibility of the \ac{CHP} plants and allows us to test different approaches for the mutate function, even though it should be noted that this behavior might not be desired or even feasible in a real scenario.  

Regarding the target schedule, we defined the requirements that a) both, \ac{CHP} units and wind power plants should be able to reach it without any additional power from the respective other unit type and b) the maximum generation from all units should not exceed the target schedule by more than the target schedule itself. 

The first requirement ensures that the Pareto front includes three "extreme" solutions with two out of three objectives having the optimal value and the last objective having the worst possible value. For example, when the target is fully reached by only wind power plants, objectives $f_1$ and $f_2$ would be $0$, while $f_3$ is $1$. This gives some information about the otherwise unknown true Pareto front and helps in the analysis of the results. The second requirement simplifies the calculation of $f_{1max}$, as it would exactly equal to twice the amount of the target schedule in every timestep.

To achieve this, the following adaptions have been made to the \ac{CHP} schedules and the wind generation forecasts:
\begin{itemize}
    \item For all \ac{CHP} units, the first schedule is defined to be the schedule with the most produced power, so all other schedules were reduced so as not to exceed this schedule in any interval. The sum of all these highest schedules is used to create the target schedule.
    \item The wind power plants should be able to reach this target schedule exactly by producing their maximum possible power in each interval. This means the sum of all wind power in each interval should not exceed the target schedule. All wind forecasts have been adapted to fulfill this requirement.
\end{itemize}

Each \ac{CHP} and each wind power plant is represented by an agent with all the necessary knowledge about available schedules or maximal power values. Agents choose schedules for their units and add them to the cluster schedule, which is then used to calculate the performance of the three objectives described in \autoref{subsec:problem_cpes}. The information necessary for the agents to compute the performance values and the hypervolume of the resulting Pareto front is passed at the beginning of the optimization. This includes the type of all units in the cluster schedule and the reference point, which is set to $(1.1, 1.1)$. 
For communication, the agents are connected in their overlay by a small-world topology, using $k=2$ \cite{strogatz2001}.

We tested MO-COHDA with this scenario by varying the pick- and mutate functions. 
\autoref{tab:settings} shows Settings A and B and their respective parameter values. Marked in yellow are those parameters that differ between the two settings. Since the \ac{CHP} plants have less flexibility with only ten schedules each, we chose simple \textit{pick} and \textit{mutate} functions for the agents to use in both settings, namely picking a random solution point and replacing the schedule there with a random other schedule. For the agents representing wind power plants, we varied picking just one solution point vs. picking all of them and used a more "intelligent" \textit{mutate} function. It uses the currently selected schedule of the solution point as a base and then randomly increases or decreases the schedule in each time interval by a random value between 0 and a maximum allowed change value. This value defines how different the mutated schedule can be from the previous schedule and was varied between a lower and a higher value in the two settings. 

\begin{table}
\caption{Overview of parameter combinations for Setting A and Setting B}
\begin{tabular}{p{0.09\textwidth}p{0.15\textwidth}p{0.15\textwidth}}
\toprule
     Parameter              & Setting A & Setting B \\
                   \midrule
pick \ac{CHP}           & pick random point             & pick random point             \\
\rowcolor{yellow}
pick wind          & pick random point             & pick all points               \\
mutate \ac{CHP}         & random schedule               & random schedule               \\
\rowcolor{yellow} 
mutate wind        & rnd neighbour $\pm 25$   & rnd neighbour $\pm 100$  \\
min change     & 0.0005                        & 0.0005                        \\
\# points & 25                            & 25                            \\
\# iterations      & 1                             & 1          \\
\bottomrule
\end{tabular}
\label{tab:settings}
\end{table}

In \ac{CPES} optimization, agents would usually run distributed on field appliances (for example, industrial Raspberry Pi / Revolution Pi) and not on the same machine. To study the behavior of MO-COHDA for distributed agents, all agents are executed on one machine\footnote{11th Gen Intel i5-1135G7 (8), Intel Device 9a49, 8910MiB / 15589MiB, 4267 MT/s,  Ubuntu 20.04.6 LTS x86\_64, 5.15.0-89-generic} as different processes run in parallel. All experiments are repeated 30 times each. Since the actual front is not known, we compare the results of the different settings with each other. 

\subsection{Results \& Discussion}
\label{subsec:results_cpes}

\begin{figure}
\centering
\includegraphics[width=\linewidth]{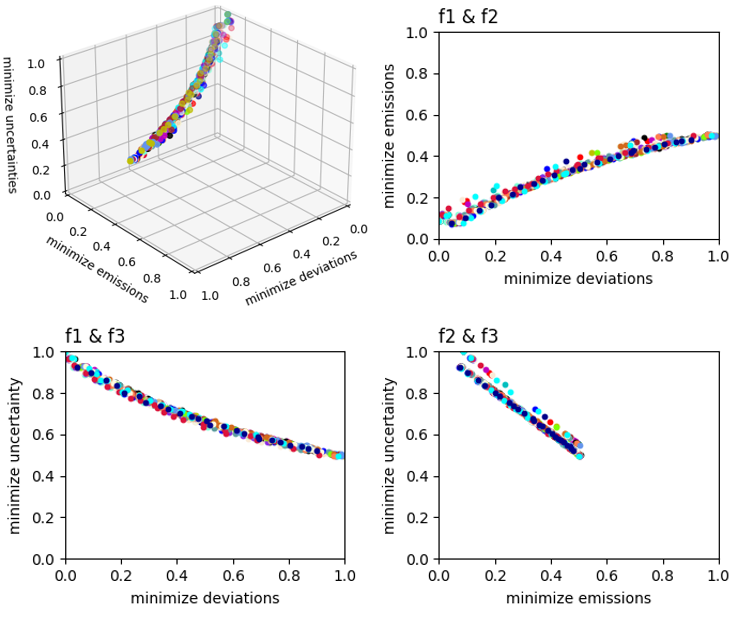}
\caption{Aggregated front of 30 runs for Setting A, executed in several processes}
\label{fig:scenario2_dis}
\Description[Four separate pictures show all the solution points found in all fronts for setting A. The top left picture shows a 3D plot with all objectives, the other pictures show 2D plots with two objectives each. The aggregated front forms a slightly curved line from high values in f1 and low values in f2 and f3 to low values in f1, low values in f2 and higher values in f3 with less points deviating from the curve]{The results for setting A are shown in four separate pictures. The top left picture shows a 3D plot with all objectives visible. The points are placed close together in the plot, nearly forming a 2D line in the 3D plot, that ranges from high values in f1 and low values in f2 and f3 to low values in f1, low values in f2 and higher values in f3. The top right picture shows the same aggregated front in a 2D plot with only objectives f1 and f2. Here, the front is not Pareto-optimal, when f1 increases, f2 also increases. The front reaches from around (0, 0) to (1.0, 0.5). Bottom left shows the 2D Plot for f1 and f3, showing a clear Pareto-front. Here the front draws a slightly curved line between (0, 1) and (1, 0.5). Finally, the bottom right 2D plot shows the front for f2 and f3, which shows a linear decrease from (0.1, 0.9) to (0.5, 0.5)}
\end{figure}

\begin{figure}
\centering
\includegraphics[width=\linewidth]{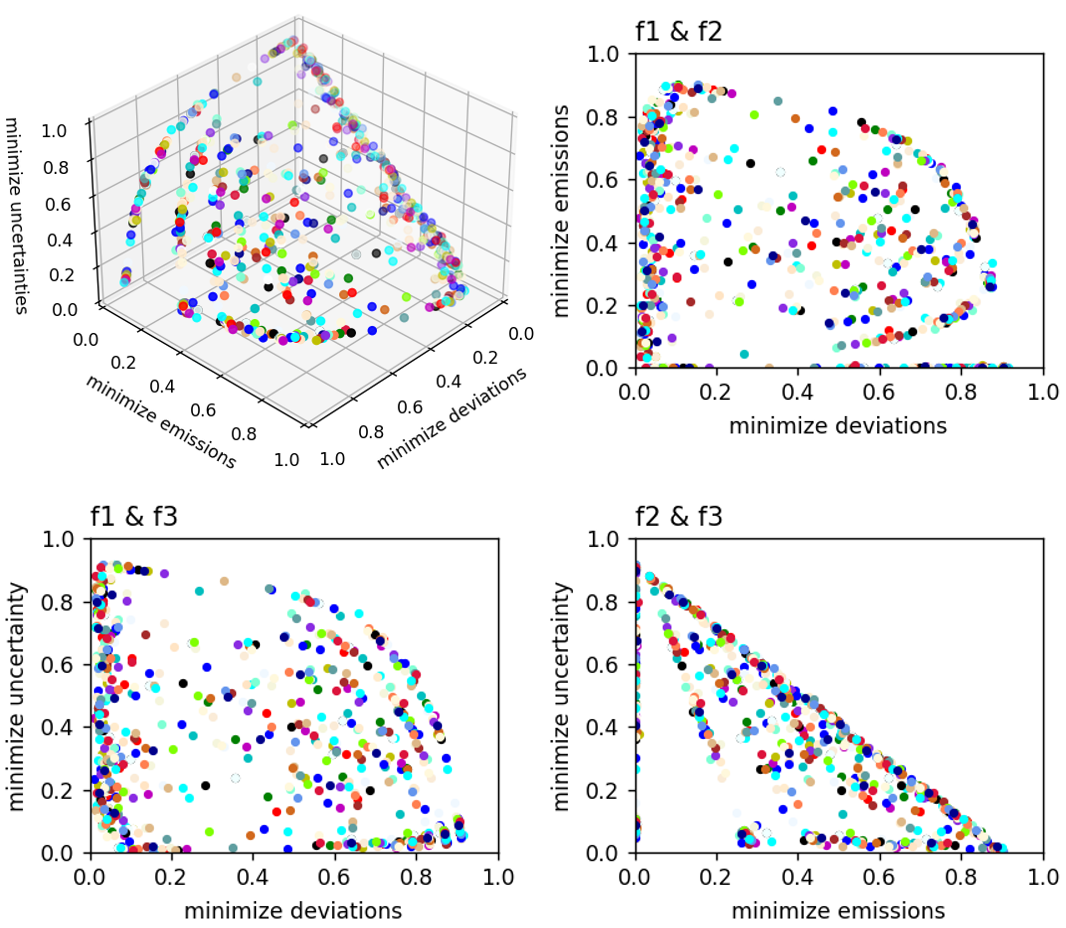}
\caption{Aggregated front of 30 runs for Setting B, executed in several processes}
\label{fig:scenario13_dis}
\Description[The aggregated front for the multi-objective optimization described in setting B is shown in four figures: including all three objectives (minimization of deviations to target, emissions and uncertainties and subfigures including two objectives each]{This figure shows the aggregated front from of in total 30 runs for Setting B of the multi-objective optimization in CPES. This setting includes differences in the pick and mutate functions. The first subfigure (top left) shows a 3D graphic, including the aggregated front, taking into account all three objectives of the optimization. In each figure, the single solutions coming from different runs are depicted in different colors. The solution points forming the front are arranged in a very distributed manner, meaning that solution points fulfilling different ranges of the respective objective functions exist. The second figure (top right) shows the results of the front regarding the first two objectives (minimization of the deviation from the target and the minimization of emissions). Here, the different solution points of the fronts are furthermore arranged very distributed. Solution points cover the range from 0.0 to 1.0 for each of the respective objectives. The third subfigure (bottom left) shows the front considering the objectives minimization of the deviations from the target and of the uncertainties. The arrangement of the points on the front is similar to the figure before, the front consists of distributed points, covering the complete ranges of the target (0.0 - 1.0) Finally, the last subfigure (bottom right) includes the front regarding the objectives for the minimization of emissions and uncertainties. The arrangement here differs from the previous ones, a clear line is recognizable, running from 0.9 of one target to 0.9 of the other target. Points that exceed 0.9 in one of the two targets are therefore not included.}
\end{figure}

\autoref{fig:scenario2_dis} and \autoref{fig:scenario13_dis} show the results from all 30 runs for setting A and setting B, respectively, depicting the entire front with three objectives and combinations of all pairs of two objectives. Each front has a different color, so it can be seen that the solution points of individual fronts are well distributed across the Pareto front. There are significant differences in the Pareto fronts of settings A and B. While the solution points for setting A in \autoref{fig:scenario2_dis} are all close together, only covering a small area of the solution space, the results for setting B in \autoref{fig:scenario13_dis} show a well-distributed front. In setting A, the performance values for minimizing emissions lie on average between $(0.08, 0.50)$ and for minimizing uncertainty between $(0.51, 0.97)$, only covering half the possible range for the performance values for those two objectives. This suggests, that in all the identified cluster schedules, the percentage of power produced by wind plants is higher than of power produced by \ac{CHP} units. Moreover, in setting A not for all combinations of objectives, Pareto-optimal solutions have been identified, as the performance of the solution points increases for both minimizing emissions and deviations. In contrast, setting B covers the full range of performance values for all three objectives, providing a greater variety of solutions to choose from.

\autoref{tab:results_cpes} shows different performance indicators with mean and standard deviation for settings A and B. HV (aggr) defines the hypervolume of the aggregated front of all 30 runs as a reference value for the mean values of individual runs. The number of decide calls is measured for the same reference agent -- a wind power plant agent -- in all scenarios. One decide call always encompasses the process described in \autoref{fig:activity_diagram}, including calculating the objective functions and the hypervolume. The number of messages considers all messages exchanged between the agents during the MO-COHDA optimization. The higher hypervolume value of setting B reflects the improved selection and distribution of solution points compared to setting A. Considering the much higher number of messages and decide-function calls, the agents optimize for a longer time, resulting in more mutations and more exploration of the solution space. When comparing the HV and HV (aggr) values for each experiment, it can also be seen that different Pareto-optimal solutions are found in the 30 runs, resulting in a slightly better HV value for the aggregated front.

\begin{table}[]
\caption{Comparing performance indicators of 30 runs of MO-COHDA for different settings and implementations (mean and standard deviation)}
\begin{tabular}{lcccc}
\toprule
& HV & \begin{tabular}[c]{@{}l@{}}HV\\ (aggr)\end{tabular} & \# decide calls & \# msgs \\
\midrule
A & $0.421 \pm 0.006$ & $0.448$    & $275 \pm 70$   & $37209 \pm 12090$    \\
B & $0.935 \pm 0.004$   & $1.012$    & $788 \pm 125$  & $230041 \pm 63065$          \\
\bottomrule
\end{tabular}
\label{tab:results_cpes}
\end{table}

While these results show that MO-COHDA can find a set of Pareto-optimal solutions in each parameter setting and implementation, they also indicate that correct parameter tuning is critical to get the best results. A trade-off between performance and computational complexity (and therefore runtime) can be observed between the two settings. Based on the concrete use case, it has to be decided what to prioritize: While some use cases might have strict time limits but do not require the optimal solution, other use cases might require the optimal solution but have more time to find it. 

\subsection{Computational Complexity of MO-COHDA}
Optimization problems in \ac{CPES} can come with certain requirements or limitations for the respective algorithm, such as a time limit to find a feasible solution or communication requirements such as a limited amount of messages. The computational complexity of optimization algorithms is therefore highly relevant in order to assess their applicability for specific problems and their scalability. This includes time complexity, space complexity (in case of limited computational power of the devices the agents are executed on), and communication complexity.

Basic COHDA has already been thoroughly analyzed regarding its scalability. The results show that with a larger number of agents, run-time increases logarithmically, and the communication expenses increase superlinearly as in $\mathcal{O}(n^{1.4})$, while increasing search spaces and planning horizons (length of the schedules) have little effect \cite{hinrichs2017distributed}. Hinrichs and Sonnenschein state, that due to the fast decision process, they were able to solve instances with thousands of agents within a few minutes.

In MO-COHDA however, the complexity of the decision process of individual agents can increase significantly, depending on the choice of parameters. The complexity of the mutate function, the number of iterations and the number of solution points to be optimized can be especially critical here. Additionally, there is not just one target function to be evaluated, but a set of target functions plus the hypervolume calculation of the whole front. In MO-COHDA we use the \textit{evoalgos}\footnote{https://www.simonwessing.de/evoalgos/doc/} package to calculate the hypervolume. This package has implemented the hypervolume calculation from Fonseca et. al which has $\mathcal{O}(n^{d-2} log n)$ time complexity and $\mathcal{O}(n)$ space complexity, with $n$ being the number of solution points and $d$ the number of objectives \cite{guerreiro2021hypervolume}.

To learn more about the scalability and limitations of MO-COHDA, further experiments are required with different parameter settings.

\section{Conclusion}
\label{sec:conclusion}
Operating a large number of \ac{DER} and other controllable elements in a \ac{CPES} requires multi-objective problem-solving. Due to the popularity of distributed algorithms for monitoring and control of \ac{CPES}, it should also be possible to perform multi-objective optimization in a fully distributed way. In this paper we introduced MO-COHDA, an agent-based optimization algorithm, which is capable of identifying a set of Pareto-optimal solutions for a given multi-objective optimization problem. MO-COHDA allows a flexible choice of optimization strategies for the different agents to best fit the respective flexibility representation, including changing these strategies over the course of the optimization. 

Using three benchmark functions from the \ac{ZDT} problem suite we showed that with the right parameter tuning MO-COHDA is capable of approximating the reference front and gives similarly good solutions as a central NSGA2 approach. We then proceeded to show the implementation of MO-COHDA in an exemplary use case from the energy domain by scheduling a set of \ac{DER} such that the distance to a target schedule, the amount of emissions and the uncertainty of the resulting schedule are minimized. By studying the results of two different parameter settings and two different forms of implementation of agents (single- and multi-processing) we could show how these different versions affect the performance.

Further research is necessary to get more insight into the parameter tuning of MO-COHDA for different multi-objective optimization problems in \ac{CPES}, especially considering the real Pareto front is usually not known. In this regard, the different agent implementations and their influence on finding optimal solutions should be studied in more detail as well.
In addition, the comparison of the presented MO-COHDA variant with other existing distributed algorithms in terms of performance, complexity, and efficiency for different \ac{CPES} use cases represents another area of future work.

\begin{acronym}
\acro{CHP}{Combined Heat and Power}
\acro{COHDA}[COHDA]{Combinatorial Optimization Heuristic for Distributed Agents}
\acro{CPES}[CPES]{Cyber-Physical Energy Systems}
\acro{DER}[DER]{Distributed Energy Resources}
\acro{MAS}[MAS]{Multi-Agent Systems}
\acro{NSGA-2}{Non-dominated Sorting Genetic Algorithm}
\acro{ICT}[ICT]{Information and Communication Technology}
\acro{ZDT}{Zitzler–Deb–Thiele}
\end{acronym}

\begin{acks}
To Jörg, for the interesting discussions about (MO-)COHDA.
\end{acks}

\bibliographystyle{ACM-Reference-Format}
\bibliography{ref}


\begin{thebibliography}{44}


\ifx \showCODEN    \undefined \def \showCODEN     #1{\unskip}     \fi
\ifx \showDOI      \undefined \def \showDOI       #1{#1}\fi
\ifx \showISBNx    \undefined \def \showISBNx     #1{\unskip}     \fi
\ifx \showISBNxiii \undefined \def \showISBNxiii  #1{\unskip}     \fi
\ifx \showISSN     \undefined \def \showISSN      #1{\unskip}     \fi
\ifx \showLCCN     \undefined \def \showLCCN      #1{\unskip}     \fi
\ifx \shownote     \undefined \def \shownote      #1{#1}          \fi
\ifx \showarticletitle \undefined \def \showarticletitle #1{#1}   \fi
\ifx \showURL      \undefined \def \showURL       {\relax}        \fi
\providecommand\bibfield[2]{#2}
\providecommand\bibinfo[2]{#2}
\providecommand\natexlab[1]{#1}
\providecommand\showeprint[2][]{arXiv:#2}

\bibitem[Adihou et~al\mbox{.}(2019)]%
        {adihou2019multi}
\bibfield{author}{\bibinfo{person}{Yolaine Adihou}, \bibinfo{person}{Mohamed~Tahar Mabrouk}, \bibinfo{person}{Pierrick Haurant}, {and} \bibinfo{person}{Bruno Lacarri{\`e}ere}.} \bibinfo{year}{2019}\natexlab{}.
\newblock \showarticletitle{A multi-objective optimization model for the operation of decentralized multi-energy systems}. In \bibinfo{booktitle}{\emph{Journal of Physics: Conference Series}}, Vol.~\bibinfo{volume}{1343}. IOP Publishing, \bibinfo{pages}{012104}.
\newblock
\urldef\tempurl%
\url{https://doi.org/10.1088/1742-6596/1343/1/012104}
\showDOI{\tempurl}


\bibitem[Almoghathawi et~al\mbox{.}(2019)]%
        {almoghathawi2019resilience}
\bibfield{author}{\bibinfo{person}{Yasser Almoghathawi}, \bibinfo{person}{Kash Barker}, {and} \bibinfo{person}{Laura~A Albert}.} \bibinfo{year}{2019}\natexlab{}.
\newblock \showarticletitle{Resilience-driven restoration model for interdependent infrastructure networks}.
\newblock \bibinfo{journal}{\emph{Reliability Engineering \& System Safety}}  \bibinfo{volume}{185} (\bibinfo{year}{2019}), \bibinfo{pages}{12--23}.
\newblock


\bibitem[Apata and Oyedokun(2020)]%
        {apata2020overview}
\bibfield{author}{\bibinfo{person}{O Apata} {and} \bibinfo{person}{DTO Oyedokun}.} \bibinfo{year}{2020}\natexlab{}.
\newblock \showarticletitle{An overview of control techniques for wind turbine systems}.
\newblock \bibinfo{journal}{\emph{Scientific African}}  \bibinfo{volume}{10} (\bibinfo{year}{2020}), \bibinfo{pages}{e00566}.
\newblock


\bibitem[Berghammer et~al\mbox{.}(2012)]%
        {berghammer2012convergence}
\bibfield{author}{\bibinfo{person}{Rudolf Berghammer}, \bibinfo{person}{Tobias Friedrich}, {and} \bibinfo{person}{Frank Neumann}.} \bibinfo{year}{2012}\natexlab{}.
\newblock \showarticletitle{Convergence of set-based multi-objective optimization, indicators and deteriorative cycles}.
\newblock \bibinfo{journal}{\emph{Theoretical Computer Science}}  \bibinfo{volume}{456} (\bibinfo{year}{2012}), \bibinfo{pages}{2--17}.
\newblock


\bibitem[Beume et~al\mbox{.}(2007)]%
        {beume2007sms}
\bibfield{author}{\bibinfo{person}{Nicola Beume}, \bibinfo{person}{Boris Naujoks}, {and} \bibinfo{person}{Michael Emmerich}.} \bibinfo{year}{2007}\natexlab{}.
\newblock \showarticletitle{SMS-EMOA: Multiobjective selection based on dominated hypervolume}.
\newblock \bibinfo{journal}{\emph{European Journal of Operational Research}} \bibinfo{volume}{181}, \bibinfo{number}{3} (\bibinfo{year}{2007}), \bibinfo{pages}{1653--1669}.
\newblock


\bibitem[Blank and Deb(2020)]%
        {blank2020pymoo}
\bibfield{author}{\bibinfo{person}{Julian Blank} {and} \bibinfo{person}{Kalyanmoy Deb}.} \bibinfo{year}{2020}\natexlab{}.
\newblock \showarticletitle{Pymoo: Multi-objective optimization in python}.
\newblock \bibinfo{journal}{\emph{Ieee access}}  \bibinfo{volume}{8} (\bibinfo{year}{2020}), \bibinfo{pages}{89497--89509}.
\newblock


\bibitem[Bremer and Lehnhoff(2017)]%
        {bremer2017agent}
\bibfield{author}{\bibinfo{person}{Joerg Bremer} {and} \bibinfo{person}{Sebastian Lehnhoff}.} \bibinfo{year}{2017}\natexlab{}.
\newblock \showarticletitle{An agent-based approach to decentralized global optimization-adapting cohda to coordinate descent}. In \bibinfo{booktitle}{\emph{International Conference on Agents and Artificial Intelligence}}, Vol.~\bibinfo{volume}{2}. SCITEPRESS, \bibinfo{pages}{129--136}.
\newblock


\bibitem[Bremer and Lehnhoff(2018)]%
        {bremer2018hybridizing}
\bibfield{author}{\bibinfo{person}{Jörg Bremer} {and} \bibinfo{person}{Sebastian Lehnhoff}.} \bibinfo{year}{2018}\natexlab{}.
\newblock \showarticletitle{Hybridizing s-metric selection and support vector decoder for constrained multi-objective energy management}. In \bibinfo{booktitle}{\emph{International Conference on Hybrid Intelligent Systems}}. Springer, \bibinfo{pages}{249--259}.
\newblock


\bibitem[Bremer and Lehnhoff(2019)]%
        {bremer2019towards}
\bibfield{author}{\bibinfo{person}{Jörg Bremer} {and} \bibinfo{person}{Sebastian Lehnhoff}.} \bibinfo{year}{2019}\natexlab{}.
\newblock \showarticletitle{Towards fully decentralized multi-objective energy scheduling}. In \bibinfo{booktitle}{\emph{2019 Federated Conference on Computer Science and Information Systems (FedCSIS)}}. IEEE, \bibinfo{pages}{193--201}.
\newblock


\bibitem[Espina et~al\mbox{.}(2020)]%
        {espina2020distributed}
\bibfield{author}{\bibinfo{person}{Enrique Espina}, \bibinfo{person}{Jacqueline Llanos}, \bibinfo{person}{Claudio Burgos-Mellado}, \bibinfo{person}{Roberto Cardenas-Dobson}, \bibinfo{person}{Manuel Martinez-Gomez}, {and} \bibinfo{person}{Doris S{\'a}ez}.} \bibinfo{year}{2020}\natexlab{}.
\newblock \showarticletitle{Distributed control strategies for microgrids: An overview}.
\newblock \bibinfo{journal}{\emph{IEEE Access}}  \bibinfo{volume}{8} (\bibinfo{year}{2020}), \bibinfo{pages}{193412--193448}.
\newblock


\bibitem[Guerreiro et~al\mbox{.}(2021)]%
        {guerreiro2021hypervolume}
\bibfield{author}{\bibinfo{person}{Andreia~P Guerreiro}, \bibinfo{person}{Carlos~M Fonseca}, {and} \bibinfo{person}{Lu{\'\i}s Paquete}.} \bibinfo{year}{2021}\natexlab{}.
\newblock \showarticletitle{The hypervolume indicator: Computational problems and algorithms}.
\newblock \bibinfo{journal}{\emph{ACM Computing Surveys (CSUR)}} \bibinfo{volume}{54}, \bibinfo{number}{6} (\bibinfo{year}{2021}), \bibinfo{pages}{1--42}.
\newblock


\bibitem[Han et~al\mbox{.}(2017)]%
        {han2017taxonomy}
\bibfield{author}{\bibinfo{person}{Xue Han}, \bibinfo{person}{Kai Heussen}, \bibinfo{person}{Oliver Gehrke}, \bibinfo{person}{Henrik~W Bindner}, {and} \bibinfo{person}{Benjamin Kroposki}.} \bibinfo{year}{2017}\natexlab{}.
\newblock \showarticletitle{Taxonomy for evaluation of distributed control strategies for distributed energy resources}.
\newblock \bibinfo{journal}{\emph{IEEE Transactions on Smart Grid}} \bibinfo{volume}{9}, \bibinfo{number}{5} (\bibinfo{year}{2017}), \bibinfo{pages}{5185--5195}.
\newblock


\bibitem[Hinrichs et~al\mbox{.}(2013)]%
        {hinrichs2013distributed}
\bibfield{author}{\bibinfo{person}{Christian Hinrichs}, \bibinfo{person}{J{\"o}rg Bremer}, {and} \bibinfo{person}{Michael Sonnenschein}.} \bibinfo{year}{2013}\natexlab{}.
\newblock \showarticletitle{Distributed hybrid constraint handling in large scale virtual power plants}. In \bibinfo{booktitle}{\emph{IEEE PES ISGT Europe 2013}}. IEEE, \bibinfo{pages}{1--5}.
\newblock
\urldef\tempurl%
\url{https://doi.org/10.1109/ISGTEurope.2013.6695312}
\showDOI{\tempurl}


\bibitem[Hinrichs and Sonnenschein(2017)]%
        {hinrichs2017distributed}
\bibfield{author}{\bibinfo{person}{Christian Hinrichs} {and} \bibinfo{person}{Michael Sonnenschein}.} \bibinfo{year}{2017}\natexlab{}.
\newblock \showarticletitle{A distributed combinatorial optimisation heuristic for the scheduling of energy resources represented by self-interested agents.}
\newblock \bibinfo{journal}{\emph{Int. J. Bio Inspired Comput.}} \bibinfo{volume}{10}, \bibinfo{number}{2} (\bibinfo{year}{2017}), \bibinfo{pages}{69--78}.
\newblock


\bibitem[Hinrichs et~al\mbox{.}(2015)]%
        {hinrichs2015self}
\bibfield{author}{\bibinfo{person}{Christian Hinrichs}, \bibinfo{person}{Michael Sonnenschein}, \bibinfo{person}{Adam Gray}, {and} \bibinfo{person}{Curran Crawford}.} \bibinfo{year}{2015}\natexlab{}.
\newblock \showarticletitle{Self-organizing demand response with comfort-constrained heat pumps}. In \bibinfo{booktitle}{\emph{EnviroInfo and ICT for Sustainability 2015}}. Atlantis Press, \bibinfo{pages}{353--360}.
\newblock
\urldef\tempurl%
\url{https://doi.org/10.2991/ict4s-env-15.2015.40}
\showDOI{\tempurl}


\bibitem[Holly and Nie{\ss}e(2021)]%
        {holly2021dynamic}
\bibfield{author}{\bibinfo{person}{Stefanie Holly} {and} \bibinfo{person}{Astrid Nie{\ss}e}.} \bibinfo{year}{2021}\natexlab{}.
\newblock \showarticletitle{Dynamic communication topologies for distributed heuristics in energy system optimization algorithms}. In \bibinfo{booktitle}{\emph{2021 16th Conference on Computer Science and Intelligence Systems (FedCSIS)}}. IEEE, \bibinfo{pages}{191--200}.
\newblock


\bibitem[Karimi and Jadid(2020)]%
        {karimi2020optimal}
\bibfield{author}{\bibinfo{person}{Hamid Karimi} {and} \bibinfo{person}{Shahram Jadid}.} \bibinfo{year}{2020}\natexlab{}.
\newblock \showarticletitle{Optimal energy management for multi-microgrid considering demand response programs: A stochastic multi-objective framework}.
\newblock \bibinfo{journal}{\emph{Energy}}  \bibinfo{volume}{195} (\bibinfo{year}{2020}), \bibinfo{pages}{116992}.
\newblock


\bibitem[Khezri and Mahmoudi(2020)]%
        {khezri2020review}
\bibfield{author}{\bibinfo{person}{Rahmat Khezri} {and} \bibinfo{person}{Amin Mahmoudi}.} \bibinfo{year}{2020}\natexlab{}.
\newblock \showarticletitle{Review on the state-of-the-art multi-objective optimisation of hybrid standalone/grid-connected energy systems}.
\newblock \bibinfo{journal}{\emph{IET Generation, Transmission \& Distribution}} \bibinfo{volume}{14}, \bibinfo{number}{20} (\bibinfo{year}{2020}), \bibinfo{pages}{4285--4300}.
\newblock


\bibitem[Khunkitti et~al\mbox{.}(2021)]%
        {khunkitti2021multi}
\bibfield{author}{\bibinfo{person}{Sirote Khunkitti}, \bibinfo{person}{Apirat Siritaratiwat}, {and} \bibinfo{person}{Suttichai Premrudeepreechacharn}.} \bibinfo{year}{2021}\natexlab{}.
\newblock \showarticletitle{Multi-objective optimal power flow problems based on slime mould algorithm}.
\newblock \bibinfo{journal}{\emph{Sustainability}} \bibinfo{volume}{13}, \bibinfo{number}{13} (\bibinfo{year}{2021}), \bibinfo{pages}{7448}.
\newblock


\bibitem[Konak et~al\mbox{.}(2006)]%
        {konak2006multi}
\bibfield{author}{\bibinfo{person}{Abdullah Konak}, \bibinfo{person}{David~W Coit}, {and} \bibinfo{person}{Alice~E Smith}.} \bibinfo{year}{2006}\natexlab{}.
\newblock \showarticletitle{Multi-objective optimization using genetic algorithms: A tutorial}.
\newblock \bibinfo{journal}{\emph{Reliability engineering \& system safety}} \bibinfo{volume}{91}, \bibinfo{number}{9} (\bibinfo{year}{2006}), \bibinfo{pages}{992--1007}.
\newblock


\bibitem[Luo et~al\mbox{.}(2018)]%
        {luo2018distributed}
\bibfield{author}{\bibinfo{person}{Fengji Luo}, \bibinfo{person}{Zhao~Yang Dong}, \bibinfo{person}{Gaoqi Liang}, \bibinfo{person}{Junichi Murata}, {and} \bibinfo{person}{Zhao Xu}.} \bibinfo{year}{2018}\natexlab{}.
\newblock \showarticletitle{A distributed electricity trading system in active distribution networks based on multi-agent coalition and blockchain}.
\newblock \bibinfo{journal}{\emph{IEEE Transactions on Power Systems}} \bibinfo{volume}{34}, \bibinfo{number}{5} (\bibinfo{year}{2018}), \bibinfo{pages}{4097--4108}.
\newblock


\bibitem[Mayer et~al\mbox{.}(2020)]%
        {mayer2020environmental}
\bibfield{author}{\bibinfo{person}{Martin~J{\'a}nos Mayer}, \bibinfo{person}{Art{\'u}r Szil{\'a}gyi}, {and} \bibinfo{person}{Gyula Gr{\'o}f}.} \bibinfo{year}{2020}\natexlab{}.
\newblock \showarticletitle{Environmental and economic multi-objective optimization of a household level hybrid renewable energy system by genetic algorithm}.
\newblock \bibinfo{journal}{\emph{Applied Energy}}  \bibinfo{volume}{269} (\bibinfo{year}{2020}), \bibinfo{pages}{115058}.
\newblock


\bibitem[Murray et~al\mbox{.}(2020)]%
        {MURRAY2020117792}
\bibfield{author}{\bibinfo{person}{Portia Murray}, \bibinfo{person}{Jan Carmeliet}, {and} \bibinfo{person}{Kristina Orehounig}.} \bibinfo{year}{2020}\natexlab{}.
\newblock \showarticletitle{Multi-Objective Optimisation of Power-to-Mobility in Decentralised Multi-Energy Systems}.
\newblock \bibinfo{journal}{\emph{Energy}}  \bibinfo{volume}{205} (\bibinfo{year}{2020}), \bibinfo{pages}{117792}.
\newblock
\showISSN{0360-5442}
\urldef\tempurl%
\url{https://doi.org/10.1016/j.energy.2020.117792}
\showDOI{\tempurl}


\bibitem[Naderi et~al\mbox{.}(2021)]%
        {naderi2021novel}
\bibfield{author}{\bibinfo{person}{Ehsan Naderi}, \bibinfo{person}{Mahdi Pourakbari-Kasmaei}, \bibinfo{person}{Fernando~V Cerna}, {and} \bibinfo{person}{Matti Lehtonen}.} \bibinfo{year}{2021}\natexlab{}.
\newblock \showarticletitle{A novel hybrid self-adaptive heuristic algorithm to handle single-and multi-objective optimal power flow problems}.
\newblock \bibinfo{journal}{\emph{International Journal of Electrical Power \& Energy Systems}}  \bibinfo{volume}{125} (\bibinfo{year}{2021}), \bibinfo{pages}{106492}.
\newblock


\bibitem[Nie{\ss}e and Tr{\"o}schel(2016)]%
        {niesse2016controlled}
\bibfield{author}{\bibinfo{person}{Astrid Nie{\ss}e} {and} \bibinfo{person}{Martin Tr{\"o}schel}.} \bibinfo{year}{2016}\natexlab{}.
\newblock \showarticletitle{Controlled self-organization in smart grids}. In \bibinfo{booktitle}{\emph{2016 IEEE International Symposium on Systems Engineering (ISSE)}}. IEEE, \bibinfo{pages}{1--6}.
\newblock
\urldef\tempurl%
\url{https://doi.org/10.1109/SysEng.2016.7753189}
\showDOI{\tempurl}


\bibitem[Nowdeh et~al\mbox{.}(2019)]%
        {nowdeh2019fuzzy}
\bibfield{author}{\bibinfo{person}{S~Arabi Nowdeh}, \bibinfo{person}{I~Faraji Davoudkhani}, \bibinfo{person}{MJ~Hadidian Moghaddam}, \bibinfo{person}{E~Seifi Najmi}, \bibinfo{person}{Almoataz~Y Abdelaziz}, \bibinfo{person}{Abdollah Ahmadi}, \bibinfo{person}{Seyed-Ehsan Razavi}, {and} \bibinfo{person}{Foad~Haidari Gandoman}.} \bibinfo{year}{2019}\natexlab{}.
\newblock \showarticletitle{Fuzzy multi-objective placement of renewable energy sources in distribution system with objective of loss reduction and reliability improvement using a novel hybrid method}.
\newblock \bibinfo{journal}{\emph{Applied Soft Computing}}  \bibinfo{volume}{77} (\bibinfo{year}{2019}), \bibinfo{pages}{761--779}.
\newblock


\bibitem[Ramli et~al\mbox{.}(2018)]%
        {ramli2018optimal}
\bibfield{author}{\bibinfo{person}{Makbul~AM Ramli}, \bibinfo{person}{HREH Bouchekara}, {and} \bibinfo{person}{Abdulsalam~S Alghamdi}.} \bibinfo{year}{2018}\natexlab{}.
\newblock \showarticletitle{Optimal sizing of PV/wind/diesel hybrid microgrid system using multi-objective self-adaptive differential evolution algorithm}.
\newblock \bibinfo{journal}{\emph{Renewable energy}}  \bibinfo{volume}{121} (\bibinfo{year}{2018}), \bibinfo{pages}{400--411}.
\newblock


\bibitem[Ren et~al\mbox{.}(2019)]%
        {ren2019multi}
\bibfield{author}{\bibinfo{person}{Fukang Ren}, \bibinfo{person}{Jiangjiang Wang}, \bibinfo{person}{Sitong Zhu}, {and} \bibinfo{person}{Yi Chen}.} \bibinfo{year}{2019}\natexlab{}.
\newblock \showarticletitle{Multi-objective optimization of combined cooling, heating and power system integrated with solar and geothermal energies}.
\newblock \bibinfo{journal}{\emph{Energy conversion and management}}  \bibinfo{volume}{197} (\bibinfo{year}{2019}), \bibinfo{pages}{111866}.
\newblock


\bibitem[Ren et~al\mbox{.}(2021)]%
        {ren2021multi}
\bibfield{author}{\bibinfo{person}{Yanhuan Ren}, \bibinfo{person}{Junqi Yu}, \bibinfo{person}{Anjun Zhao}, \bibinfo{person}{Wenqiang Jing}, \bibinfo{person}{Tong Ran}, {and} \bibinfo{person}{Xiong Yang}.} \bibinfo{year}{2021}\natexlab{}.
\newblock \showarticletitle{A multi-objective operation strategy optimization for ice storage systems based on decentralized control structure}.
\newblock \bibinfo{journal}{\emph{Building Services Engineering Research and Technology}} \bibinfo{volume}{42}, \bibinfo{number}{1} (\bibinfo{year}{2021}), \bibinfo{pages}{62--81}.
\newblock
\urldef\tempurl%
\url{https://doi.org/10.1177/0143624420966259}
\showDOI{\tempurl}


\bibitem[Ruiming(2019)]%
        {ruiming2019multi}
\bibfield{author}{\bibinfo{person}{Fang Ruiming}.} \bibinfo{year}{2019}\natexlab{}.
\newblock \showarticletitle{Multi-objective optimized operation of integrated energy system with hydrogen storage}.
\newblock \bibinfo{journal}{\emph{International Journal of Hydrogen Energy}} \bibinfo{volume}{44}, \bibinfo{number}{56} (\bibinfo{year}{2019}), \bibinfo{pages}{29409--29417}.
\newblock


\bibitem[Salkuti(2019)]%
        {salkuti2019optimal}
\bibfield{author}{\bibinfo{person}{Surender~Reddy Salkuti}.} \bibinfo{year}{2019}\natexlab{}.
\newblock \showarticletitle{Optimal power flow using multi-objective glowworm swarm optimization algorithm in a wind energy integrated power system}.
\newblock \bibinfo{journal}{\emph{International Journal of Green Energy}} \bibinfo{volume}{16}, \bibinfo{number}{15} (\bibinfo{year}{2019}), \bibinfo{pages}{1547--1561}.
\newblock


\bibitem[Schrage et~al\mbox{.}(2023a)]%
        {schrage2023mango}
\bibfield{author}{\bibinfo{person}{Rico Schrage}, \bibinfo{person}{Jens Sager}, \bibinfo{person}{Jan~Philipp Hörding}, {and} \bibinfo{person}{Stefanie Holly}.} \bibinfo{year}{2023}\natexlab{a}.
\newblock \bibinfo{title}{mango: A Modular Python-Based Agent Simulation Framework}.
\newblock
\newblock
\showeprint[arxiv]{2311.17688}~[cs.MA]


\bibitem[Schrage et~al\mbox{.}(2023b)]%
        {schrage2023multi}
\bibfield{author}{\bibinfo{person}{Rico Schrage}, \bibinfo{person}{Paul~Hendrik Tiemann}, {and} \bibinfo{person}{Astrid Nie{\ss}e}.} \bibinfo{year}{2023}\natexlab{b}.
\newblock \showarticletitle{A Multi-Criteria Metaheuristic Algorithm for Distributed Optimization of Electric Energy Storage}.
\newblock \bibinfo{journal}{\emph{ACM SIGENERGY Energy Informatics Review}} \bibinfo{volume}{2}, \bibinfo{number}{4} (\bibinfo{year}{2023}), \bibinfo{pages}{44--59}.
\newblock


\bibitem[Sharma et~al\mbox{.}(2020)]%
        {sharma2020coordination}
\bibfield{author}{\bibinfo{person}{Sachin Sharma}, \bibinfo{person}{KR Niazi}, \bibinfo{person}{Kusum Verma}, {and} \bibinfo{person}{Tanuj Rawat}.} \bibinfo{year}{2020}\natexlab{}.
\newblock \showarticletitle{Coordination of different DGs, BESS and demand response for multi-objective optimization of distribution network with special reference to Indian power sector}.
\newblock \bibinfo{journal}{\emph{International Journal of Electrical Power \& Energy Systems}}  \bibinfo{volume}{121} (\bibinfo{year}{2020}), \bibinfo{pages}{106074}.
\newblock


\bibitem[Stark et~al\mbox{.}(2021)]%
        {stark2021your}
\bibfield{author}{\bibinfo{person}{Sanja Stark}, \bibinfo{person}{Anna Volkova}, \bibinfo{person}{Sebastian Lehnhoff}, {and} \bibinfo{person}{Hermann de Meer}.} \bibinfo{year}{2021}\natexlab{}.
\newblock \showarticletitle{Why your power system restoration does not work and what the ICT system can do about it}. In \bibinfo{booktitle}{\emph{Proceedings of the twelfth ACM international conference on future energy systems}}. \bibinfo{address}{Virtual Event, Italy}, \bibinfo{pages}{269--273}.
\newblock


\bibitem[Strogatz(2001)]%
        {strogatz2001}
\bibfield{author}{\bibinfo{person}{Steven Strogatz}.} \bibinfo{year}{2001}\natexlab{}.
\newblock \showarticletitle{Strogatz, S.H.: Exploring Complex Networks. Nature 410, 268}.
\newblock \bibinfo{journal}{\emph{Nature}}  \bibinfo{volume}{410} (\bibinfo{date}{04} \bibinfo{year}{2001}), \bibinfo{pages}{268--76}.
\newblock
\urldef\tempurl%
\url{https://doi.org/10.1038/35065725}
\showDOI{\tempurl}


\bibitem[Ullah et~al\mbox{.}(2021)]%
        {ullah2021multi}
\bibfield{author}{\bibinfo{person}{Kalim Ullah}, \bibinfo{person}{Ghulam Hafeez}, \bibinfo{person}{Imran Khan}, \bibinfo{person}{Sadaqat Jan}, {and} \bibinfo{person}{Nadeem Javaid}.} \bibinfo{year}{2021}\natexlab{}.
\newblock \showarticletitle{A multi-objective energy optimization in smart grid with high penetration of renewable energy sources}.
\newblock \bibinfo{journal}{\emph{Applied Energy}}  \bibinfo{volume}{299} (\bibinfo{year}{2021}), \bibinfo{pages}{117104}.
\newblock


\bibitem[Wang et~al\mbox{.}(2020)]%
        {wang2020multi}
\bibfield{author}{\bibinfo{person}{Rui Wang}, \bibinfo{person}{Jian Xiong}, \bibinfo{person}{Min-fan He}, \bibinfo{person}{Liang Gao}, {and} \bibinfo{person}{Ling Wang}.} \bibinfo{year}{2020}\natexlab{}.
\newblock \showarticletitle{Multi-objective optimal design of hybrid renewable energy system under multiple scenarios}.
\newblock \bibinfo{journal}{\emph{Renewable Energy}}  \bibinfo{volume}{151} (\bibinfo{year}{2020}), \bibinfo{pages}{226--237}.
\newblock


\bibitem[Wang et~al\mbox{.}(2021)]%
        {wang2021decentralized}
\bibfield{author}{\bibinfo{person}{Xiaoxue Wang}, \bibinfo{person}{Liting Gu}, {and} \bibinfo{person}{Dong Liang}.} \bibinfo{year}{2021}\natexlab{}.
\newblock \showarticletitle{Decentralized and Multi-Objective Coordinated Optimization of Hybrid AC/DC Flexible Distribution Networks}.
\newblock \bibinfo{journal}{\emph{Frontiers in Energy Research}} (\bibinfo{year}{2021}), \bibinfo{pages}{646}.
\newblock
\urldef\tempurl%
\url{https://doi.org/10.3389/fenrg.2021.762423}
\showDOI{\tempurl}


\bibitem[Wang et~al\mbox{.}(2022)]%
        {WANG20223063}
\bibfield{author}{\bibinfo{person}{Yongli Wang}, \bibinfo{person}{Lu Guo}, \bibinfo{person}{Yuze Ma}, \bibinfo{person}{Xu Han}, \bibinfo{person}{Juntai Xing}, \bibinfo{person}{WenQiang Miao}, {and} \bibinfo{person}{Huan Wang}.} \bibinfo{year}{2022}\natexlab{}.
\newblock \showarticletitle{Study on operation optimization of decentralized integrated energy system in northern rural areas based on multi-objective}.
\newblock \bibinfo{journal}{\emph{Energy Reports}}  \bibinfo{volume}{8} (\bibinfo{year}{2022}), \bibinfo{pages}{3063--3084}.
\newblock
\showISSN{2352-4847}
\urldef\tempurl%
\url{https://doi.org/10.1016/j.egyr.2022.01.246}
\showDOI{\tempurl}


\bibitem[Wu et~al\mbox{.}(2020)]%
        {wu2020novel}
\bibfield{author}{\bibinfo{person}{Chunying Wu}, \bibinfo{person}{Jianzhou Wang}, \bibinfo{person}{Xuejun Chen}, \bibinfo{person}{Pei Du}, {and} \bibinfo{person}{Wendong Yang}.} \bibinfo{year}{2020}\natexlab{}.
\newblock \showarticletitle{A novel hybrid system based on multi-objective optimization for wind speed forecasting}.
\newblock \bibinfo{journal}{\emph{Renewable energy}}  \bibinfo{volume}{146} (\bibinfo{year}{2020}), \bibinfo{pages}{149--165}.
\newblock


\bibitem[Yang et~al\mbox{.}(2021)]%
        {yang2021multiagent}
\bibfield{author}{\bibinfo{person}{Lun Yang}, \bibinfo{person}{Yinliang Xu}, \bibinfo{person}{Hongbin Sun}, \bibinfo{person}{Moyuen Chow}, {and} \bibinfo{person}{Jianguo Zhou}.} \bibinfo{year}{2021}\natexlab{}.
\newblock \showarticletitle{A multiagent system based optimal load restoration strategy in distribution systems}.
\newblock \bibinfo{journal}{\emph{International Journal of Electrical Power \& Energy Systems}}  \bibinfo{volume}{124} (\bibinfo{year}{2021}), \bibinfo{pages}{106314}.
\newblock


\bibitem[Zitzler et~al\mbox{.}(2000)]%
        {zitzler2000comparison}
\bibfield{author}{\bibinfo{person}{Eckart Zitzler}, \bibinfo{person}{Kalyanmoy Deb}, {and} \bibinfo{person}{Lothar Thiele}.} \bibinfo{year}{2000}\natexlab{}.
\newblock \showarticletitle{Comparison of multiobjective evolutionary algorithms: Empirical results}.
\newblock \bibinfo{journal}{\emph{Evolutionary computation}} \bibinfo{volume}{8}, \bibinfo{number}{2} (\bibinfo{year}{2000}), \bibinfo{pages}{173--195}.
\newblock


\bibitem[Zitzler and K{\"u}nzli(2004)]%
        {zitzler2004indicator}
\bibfield{author}{\bibinfo{person}{Eckart Zitzler} {and} \bibinfo{person}{Simon K{\"u}nzli}.} \bibinfo{year}{2004}\natexlab{}.
\newblock \showarticletitle{Indicator-based selection in multiobjective search}. In \bibinfo{booktitle}{\emph{International conference on parallel problem solving from nature}}. Springer, \bibinfo{pages}{832--842}.
\newblock


\end{thebibliography}

\appendix

\section{Example for non-transitivity of reference point}
\label{apdx:ref_point}

\begin{figure}[H]
\centering
\includegraphics[width=0.75\linewidth]{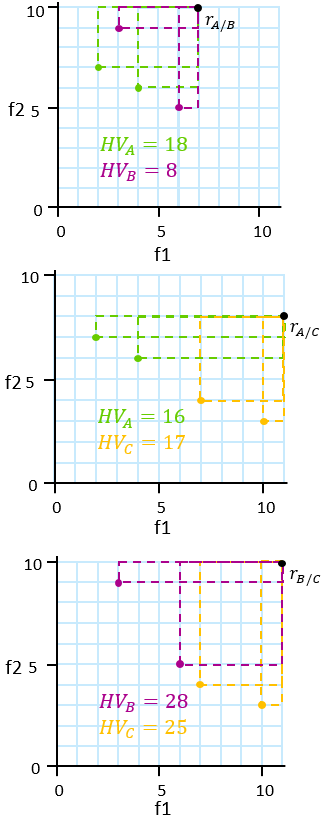}
\caption{Example for non-transitivity with dynamic reference point}
\label{fig:ref_point}
\Description[The figure shows three images. In each image 2 out of 3 fronts are shown together with two solution points each. The hypervolume between each front and the reference point is shown. The HV values per front are different, depending on the reference point. Depending on the sequence in which the fronts are compared, a different front would be considered as the best]{Three graphs are shown. In each graph two out of a total of three fronts are shown. The fronts are named A, B and C and have two solution points each. The first graph shows fronts A and B and the reference point calculated for A and B. The HV for A is 18 and for B is 8. The second graph shows fronts A and C, with the respective reference point. Here, the HV for A is 16 and for C 17. The third graph shows fronts B and C, with the HV of B being 28 and of C being 25.}
\end{figure}

Figure \ref{fig:ref_point} shows a simple example of non-transitivity of hypervolume with changing reference points. Three different Pareto fronts for minimization problems A, B and C are compared. The reference point for comparing two fronts is chosen using according to the SMS-EMOA approach (worst objective values increased by 1.0). When looking at the resulting hypervolumes we can see the deteriorative cycle with $A > B$, $B > C$ and $C > A$.

\section{Results Benchmark}
\label{apdx:benchmark}

\begin{figure}[H]
\centering
\includegraphics[width=\linewidth]{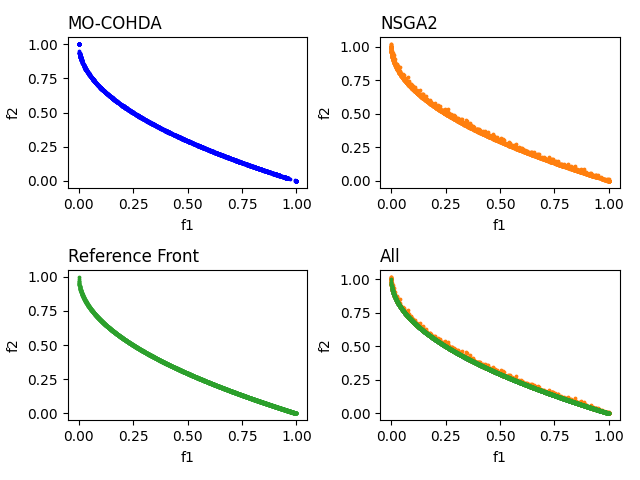}
\caption{Aggregated front of 100 runs each for MO-COHDA and Central approach compared with reference front for ZDT1}
\label{fig:zdt1}
\Description[Four images show four graphs for MO-COHDA, NSGA-2, Reference front and All together. The fronts shown in all the pictures are nearly the same, except for a few outliers in the NSGA-2 front.]{The image is distributed in four graphs, each showing the front for ZDT1 created by MO-COHDA (top left), NSGA2 (top right), reference front (bottom left) and all fronts combined (bottom right). The reference front shows the true front for ZDT1 which is a convex front reaching from (0,1) to (1,0). The fronts for MO-COHDA and NSGA-2 look nearly identical to the reference front, only NSGA-2 has a few outliers. These are especially visible in the picture showing all fronts together, where the reference front does not cover all of the points found by NSGA-2.}
\end{figure}

\begin{figure}[H]
\centering
\includegraphics[width=\linewidth]{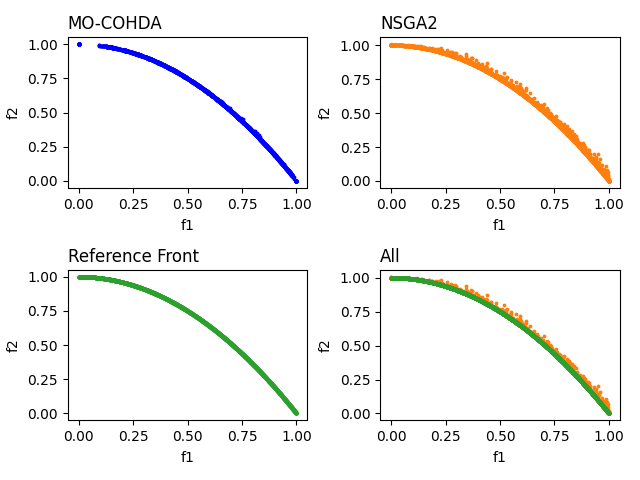}
\caption{Aggregated front of 100 runs each for MO-COHDA and Central approach compared with reference front for ZDT2}
\label{fig:zdt2}
\Description[Four images show four graphs for MO-COHDA, NSGA-2, Reference front and All together. The fronts shown in all the pictures are nearly the same, except for a few outliers in the NSGA-2 front.]{The image is distributed in four graphs, each showing the front for ZDT2 created by MO-COHDA (top left), NSGA2 (top right), reference front (bottom left) and all fronts combined (bottom right). The reference front shows the true front for ZDT2 which is a non-convex front that reaches from (0, 1) to (1,0). The fronts for MO-COHDA and NSGA-2 look nearly identical to the reference front, only NSGA-2 has a few outliers. These are especially visible in the picture showing all fronts together, where the reference front does not cover all of the points found by NSGA-2.}
\end{figure}

\end{document}